\documentstyle[aps,preprint]{revtex}

\begin{document}

\title{Monte Carlo Simulation of the Heisenberg Antiferromagnet on a
Triangular Lattice: Topological Excitations}

\author{M.\ Wintel and H.\ U.\ Everts}
\address{Institut f\"ur Theoretische Physik, Universit\"at Hannover\\
Appelstr.\ 2, D-30167 Hannover, Germany}

\author{W.\ Apel}
\address{Physikalisch-Technische Bundesanstalt\\
Bundesallee 100, D-38116 Braunschweig, Germany}

\maketitle

\begin{abstract}
We have simulated the classical Heisenberg antiferromagnet on a
triangular lattice using a local Monte Carlo algorithm.
The behavior of the correlation length $\xi$, the susceptibility at
the ordering wavevector $\chi({\bf Q})$, and the spin stiffness $\rho$
clearly reflects the existence of two temperature regimes -- a high
temperature regime $T > T_{th}$, in which the disordering effect of
vortices is dominant, and a low temperature regime $T < T_{th}$, where
correlations are controlled by small amplitude spin fluctuations.
As has previously been shown, in the last regime, the behavior of the
above quantities agrees well with the predictions of a renormalization
group treatment of the appropriate nonlinear sigma model.
For $T > T_{th}$, a satisfactory fit of the data is achieved, if the
temperature dependence of $\xi$ and $\chi({\bf Q})$ is assumed to be of
the form predicted by the Kosterlitz--Thouless theory.
Surprisingly, the crossover between the two regimes appears to happen
in a very narrow temperature interval around $T_{th} \simeq 0.28$.\\[0.5cm]
\end{abstract}


\thispagestyle{empty}

\newpage
\section{Introduction}

Magnetic ordering phenomena of both classical and quantum
antiferromagnets on non--bipartite lattices are a fascinating subject.
The simplest and most frequently studied model of this type is the
Heisenberg antiferromagnet on a triangular lattice (HAFT).  While the
question of whether or not the typical features of this model have
been observed in experiments is still a controversial issue \cite{EXP90},
the theoretical understanding of these features has advanced rapidly
during recent years
\cite{ADJ90,ADM92,AWE92,LR93,DE93}. In contrast to antiferromagnets on
bipartite lattices, the HAFT exhibits non--collinear magnetic order in
its classical and most likely also in its quantum ground state.  As a
consequence, the order parameter of the HAFT is represented locally by
a set of three mutually orthogonal unit vectors or, alternatively, by
a rotation matrix which defines the local orientation of this set
relative to some fixed frame of reference. Renormalization group (RG)
studies of appropriate nonlinear sigma (NL$\sigma$) models
\cite{ADM92,AWE92} have revealed a number of interesting properties of
the HAFT.
The symmetry of the model was found to be dynamically
enhanced from $O(3)\otimes O(2)$ to $O(4)$, and in a two loop RG
calculation for the classical HAFT \cite{ADM92}, the temperature
dependence of the correlation length $\xi$ was obtained as
\begin{equation}
\xi  = \Delta \; C^{\xi}_{RG} \; \sqrt{T/B} \; e^{B/T}\; ,
\label{1}
\end{equation}
where $\Delta$ is the lattice constant,
$B=\sqrt{3} \pi (\frac{\pi}{4} + \frac{1}{2}) =6.994$. The prefactor
$C^{\xi}_{RG}$ is left undetermined by the RG calculation.

It follows from topological considerations that the order--parameter
field of the HAFT allows for excitations of the form of $Z_2$ vortices
\cite{KM84,M79}. A numerical study of the classical HAFT \cite{KM84} has
revealed that these vortices become abundant above a threshold
temperature $T_{th} \simeq 0.3$ and that they unbind for $T>T_{th}$,
similarly as the $Z$ vortices in the planar $XY$ model above the
Kosterlitz--Thouless (KT) transition temperature $T_{KT}$ \cite{KT73}.
Further evidence for this similarity beween the dissociation mechanism
of the $Z_2$ vortices and that of the $Z$ vortices has been provided by
Kawamura and Kikuchi \cite{KK93}.
In recent work \cite{WEA94}, we have studied the influence of the
vortices on the partition function of the classical HAFT on the basis
of the NL$\sigma$ model. While a true KT--type phase--transition can
be ruled out for the HAFT, our results suggest that for $T>T_{th}$
the vortices will affect the properties of the HAFT rather drastically.
In particular, the disorder induced by the unbinding of the vortices
can be expected to lead to a crossover from the T--dependence Eq. (\ref{1})
of the correlation length in the low temperature regime $T<T_{th}$
to a KT--type behavior
\begin{equation}
  \xi = \Delta \; C^{\xi}_{KT} \; \exp( b/(T-T_{th})^{\frac{1}{2}} )
\label{2}
\end{equation}
in the high temperature regime.

It is the aim of the present paper to supplement our recent analytical
study \cite{WEA94}, which was based on a continuum description of the
HAFT by a Monte Carlo (MC) simulation of the original lattice model. A
similar study has recently been published by Southern and Young
\cite{SY93}. While we closely follow the method of these authors, our
conclusions will be quite different from theirs.

In the next Section, we first give a brief account of the technique we
used. Subsequently we present the results for the correlation length
$\xi$ and the antiferromagnetic susceptibility $\chi({\bf Q})$. While
these quantities are directly accessible to simulations in the high
temperature regime, where the disordering effect of the vortices
limits the range of the correlations, the key quantity to be computed
in the low temperature regime is the spin stiffness
\cite{CADM94,SY93}. Our numerical results for this last quantity will
be presented and discussed in Section III. Finally, we summarize the
evidence for a vortex induced crossover transition in Section IV.

\section{Correlation Length and antiferromagnetic Susceptibility}

The classsical Hamiltonian of the triangular Heisenberg
antiferromagnet can be defined as
\begin{equation}
H =  \sum_{<i,j>} {\bf S}_i \cdot {\bf S}_j \;\;.
\label{3}
\end{equation}
Here, the ${\bf S}_i$ are three dimensional unit vectors and the sum
extends over all distinct pairs of nearest neighbor sites of a
triangular lattice of $L^{2}$ sites. The exchange constant has been
set to unity. The classical ground state of the
Hamiltonian Eq. (\ref{3}) is a coplanar arrangement in which the spins on
the three sublattices are oriented at $120^{o}$ relative to each other,
\begin{equation}
 {\bf S}_i = {\bf e}_{1} cos( {\bf Q} {\bf R}_{i} ) +
          {\bf e}_{2} sin( {\bf Q} {\bf R}_{i} )\;\;.
\label{4}
\end{equation}
Here, ${\bf e}_{1}$, ${\bf e}_{2}$ are a pair of mutually orthogonal
unit vectors and ${\bf Q}$ can be any one of the six vectors pointing
towards the corners of the hexagonal Brillouin zone of the triangular
lattice, e.g. ${\bf Q} = \frac{2\pi}{\Delta} (\frac{2}{3},0)$.  The
correlation length $\xi$ can be obtained assuming a Ornstein Zernicke
form for the structure factor
\begin{eqnarray}
S({\bf q}) &=& \frac{1}{L^{2}} \sum_{i,j}
              e^{i {\bf q} ({\bf R}_{i} - {\bf R}_{j}) }
              < {\bf S}_i \cdot {\bf S}_j >           \label{5}\\
         &=& \frac{S({\bf Q})} {1+ \xi^{2} ({\bf q}-{\bf Q})^{2} }\;\;.
\nonumber
\end{eqnarray}
$\chi({\bf Q})=S({\bf Q})/T$ is then the susceptibility of the system at
the ordering wavevector.  The spin correlations $< {\bf S}_i \cdot
{\bf S}_j >$ in Eq. (\ref{5}) can be determined by MC techniques.
We used the local algorithm  described by Kawamura and Miyashita \cite{KM84}.
The lattice is divided into independent sublattices.
Then, the spins of each of these sublattices are updated sequentially.
For a given spin, a new
direction is chosen at random and the standard Metropolis rule is used
to decide whether the new direction is to be accepted.  If it is
discarded, a precessional motion through a randomly chosen angle about
the direction of the local mean field is performed.  We apply this
method to systems of linear sizes $L=12 \cdot 2^{n}, n=0,1,..5$. For
the smaller systems, $n\le3$, we discard the first $2\cdot10^4$ sweeps
for equilibration and average over the next $2\cdot 10^5$ sweeps.  For
$n=4$, we average over $4\cdot 10^5$ sweeps after discarding the
initial $10^5$ sweeps, and for $n=5$, the average is over $1.8\cdot
10^6$, and $2\cdot 10^5$ sweeps are discarded.

A selection of results for the correlation length $\xi$ and for the
antiferromagnetic susceptibility $\chi({\bf Q})$ which have been
obtained by averaging over 3-5 independent runs of these lengths is
tabulated in Table \ref{tab_ksi_chi}. As will become apparent shortly,
the data shown in this Table are crucial in checks of theoretical
predictions for the temperature dependence of $\xi$ and $\chi({\bf Q})$.
To exhibit possible finite size effects, the Table contains two pairs
of data for each temperature which correspond to two different system
sizes $L$, $2L$. In general, $\xi$ and $\chi({\bf Q})$ decrease with $T$, but
increase with the system size $L$. If, for a given temperature $T_0$
and a given system size $L_0$, the data for $\xi$ and $\chi({\bf Q})$ exhibit
no size dependence upon doubling the system size, then one can conclude
that the system size $L_0$ suffices to obtain size independent data for
all $T>T_0$. Data which are size independent by this criterion are
marked by an asterisk in Table \ref{tab_ksi_chi}. Obviously, we cannot
exclude that the data for the lowest temperature $T=0.3$ obtained
for the $L=384$ system are still size dependent. Certainly, however,
our data for $T=0.3$ are lower bounds to the thermodynamic limits of
$\xi$ and $\chi({\bf Q})$ at this temperature. To facilitate the comparison
of our data with the RG predictions we include in Table \ref{tab_ksi_chi}
the values for $\xi$ and $\chi({\bf Q})$ which result from fits of the
expressions Eq. (\ref{1}) and Eq. (\ref{6}) to these data \cite{SY93}.

In Figs. \ref{ksi-T} and \ref{chi-T}, we show our complete sets of
results for the correlation length and for the antiferromagnetic
susceptibility as functions of  the temperature. Obviously, for any
given system size $L$, there is an inflection point in the sequences
of data for $\xi$ and $\chi({\bf Q})$. This point defines a temperature
$T_L$ below which both $\xi$ and $\chi({\bf Q})$ begin to exhibit finite
size effects. In fact, as can be seen in Fig. \ref{ksi-T}, the
correlation length increases linearly with the system size for
sufficiently low temperatures $T \ll T_L$.
Figs. \ref{ksi-T} and \ref{chi-T} also
contain fits of different theoretical predictions to the numerical
data. The dashed lines represent fits of the RG result, Eq. (\ref{1}),
to our data for $\xi$ and of the form
\begin{equation}
   \chi({\bf Q})=C^{\chi}_{RG}(T/B)^4\exp(2B/T) \;,
\label{6}
\end{equation}
proposed by Southern and Young \cite{SY93} on the basis of RG
calculations, to our data for $\chi({\bf Q})$.
In these RG predictions, the constants $C^{\xi}_{RG}$ and
$C^{\chi}_{RG}$ are the only undetermined parameters. In our fits,
we neglect the data points for temperatures $T \leq T_L$ which
contain finite size effects. In agreement with Southern and Young
\cite{SY93} we find $C^{\xi}_{RG} \simeq 3 \cdot 10^{-8}$ and
$C^{\chi}_{RG}\simeq 6 \cdot 10^{-12}$. The solid lines represent
fits of the KT forms Eq. (\ref{2}) and
\begin{equation}
  \chi({\bf Q})=C^{\chi}_{KT}
  \exp( \frac{7}{4} \cdot b/(T-T_{th})^{\frac{1}{2}} )
\label{7}
\end{equation}
to the data.  With the KT form for $\xi$, Eq. (\ref{2}), the last
expression follows from the general relation
$S({\bf Q})\sim \xi^{2-\eta}$ for the structure factor
at the ordering wave vector, if $\eta$ is assumed to take the value
$\eta=1/4$ as for a proper KT transition. For the threshold
temperature, we use the value $T_{th}=0.28$, which we can be inferred
from the temperature dependence of the spin stiffness as will be
discussed in the next section. This leaves the constant $b$ which is
common to the expressions Eq. (\ref{2}) and Eq. (\ref{7}) and the
constant factors $C^{\xi}_{KT}$ and $C^{\chi}_{KT}$ of
Eq. (\ref{2}) and Eq. (\ref{7}) as fit parameters.
As before, we ignore data points for $T \leq T_L$ in our
fits. The results are $b=0.77$, $C^{\xi}_{KT}=0.47$ and
$C^{\chi}_{KT}=0.40$. Obviously, for temperatures $T\geq0.3$, the
$\exp(b/(T-T_{th})^{\frac{1}{2}})$-temperature dependence of the KT
forms fits the data better than the $\exp(B/T)$-temperature dependence
predicted by the RG analysis. In Figs. \ref{ksi-wT} and \ref{chi-wT}
we plot $\xi$ and $\chi({\bf Q})$ logarithmically against
$(T-T_{th})^{-1/2}$ so that the KT forms Eq. (\ref{2}) and
Eq. (\ref{7}) appear as straight lines.
These lines are seen to fit the data quite well in the
temperature interval $0.30 \leq T \leq 0.34$, whereas the agreement
between the curves representing the RG forms is restricted to a narrow
interval around $T \approx 0.31$. In particular, we emphasize that for
$T=0.3$, the RG predictions are incompatible with the data points for
$\xi$ and $\chi({\bf Q})$ which are listed in Table \ref{tab_ksi_chi} with
their respective errors.
In this context, we recall that if our data for $T=0.3$ do not represent
the thermodynamic limits of $\xi$ and $\chi({\bf Q})$, they are certainly
lower bounds to these limits. Hence, the discrepancy between the RG
predictions and the true values of $\xi$ and $\chi({\bf Q})$ may even
be larger than has been inferred here.
We note that the fit of the KT form, Eq. (\ref{7}),
to the data for the susceptibility $\chi({\bf Q})$ is better than that
of the RG form, Eq. (\ref{2}), to $\xi$. This may be attributable to the
lower quality of the data for $\xi$ which are obtained indirectly from
the Ornstein-Zernicke expression Eq. (\ref{5}) in the limit
$\xi|{\bf Q}- {\bf q}|\ll 1$.

In summary, we observe that our results combined with the earlier
findings of Kawamura and Miyashita \cite{KM84} support the view that
in the temperature range $T>T_{th}$ the spin correlations of the HAFT
are decisively influenced by unbound vortices, so that a perturbative
treatment of the model is inadequate in this temperature regime. It
should be obvious, however, that in the above analysis of our numerical
results we have been guided by our previous prediction \cite{WEA94} that,
in the case of the HAFT, the vortex unbinding mechnism leads to a KT-type
temperature dependence of the correlation length above a crossover
temperature $T_{th}$. While we do not claim to have found compelling
evidence for this prediction, we regard our numerical results as
strong support for it.

\section{spin stiffness}

To further corroborate the above view and in order to get insight into
the low temperature regime, where the correlation length exceeds the
accessible system sizes, we also determined the spin stiffness $\rho$
in our simulations.

The diagonal components $\rho_{\alpha}, \alpha = 1,2,3$, of the spin
stiffness tensor are the second derivatives of the free energy density
$f(\theta_{\alpha})$ with respect to the twist angles
$\theta_{\alpha}$ of the spins around three mutually orthogonal axes
${\bf e}_{\alpha}$ \cite{CADM94,SY93},
\begin{eqnarray}
     \rho_{\alpha} =
 &-& \frac{2}{\sqrt{3} L^2}  < \sum_{<i,j>}
      \left[ {\bf S}_i \cdot {\bf S}_j -
      ({\bf S}_i \cdot {\bf e}_{\alpha})
      ({\bf S}_j \cdot {\bf e}_{\alpha}) \right]
      ({\bf u} \cdot {\bf e}_{ij})^2   >\nonumber \\
 &-& \frac{2}{\sqrt{3} T L^2}  <  \left[ \sum_{<i,j>}
          {\bf S}_i \times {\bf S}_j \cdot {\bf e}_{\alpha}
           ({\bf u} \cdot {\bf e}_{ij}) \right]^2 >  \;\; .
\label{8}
\end{eqnarray}
Here, ${\bf u}$ is the lattice direction along which the twist is
applied and ${\bf e}_{ij}$ is the direction of the bond between
nearest neighbor lattice sites $i$ and $j$. The prefactor in Eq. (\ref{8})
has been chosen such that Eq. (\ref{8}) is the stiffness per unit area.

In the simulation, the thermal averages on the right hand side of
Eq. (\ref{8}) are replaced by averages over configurations which are
generated by a large number of successive MC sweeps.  For a finite
system, the ordered spin structure will change its orientation in
space in the course of the simulation. In a sufficiently large number
of sweeps, one will therefore measure the average spin stiffness
\begin{equation}
   \rho = \frac{1}{3} \sum_{\alpha=1}^{3} \rho_{\alpha} \;\;.
\label{9}
\end{equation}

Since there is no long range order in the HAFT for any finite temperature,
$\rho$ must vanish for the HAFT in the thermodynamic limit for any finite
temperature. However, for finite system sizes $L$, $\rho$ will be finite
for sufficiently low temperatures such that $\xi >L$.
According to Eq. (\ref{1}), this condition should be satisfied for system sizes
$L<10^3$ up to temperatures of the order of unity, unless the constant
$C^{\xi}_{RG}$ is exceedingly small as has been suggested by Southern
and Young \cite{SY93}.
As we have argued above, the results for $\xi $ in the high temperature
regime, $T>T_{th}$, should not be interpreted as evidence for such a
small value of $C^{\xi}_{RG}$. In fact, a naive integration of the 2--loop
renormalization group equations which starts with the microscopic
parameters of the HAFT as initial values yields the result Eq. (\ref{1}) with
$C^{\xi}_{RG}=\sqrt{(\frac{\pi}{4}+\frac{1}{2})}
 e^{- (\frac{\pi}{4}+\frac{1}{2})} = 0.314$.

In our simulations, we determine the three $\rho_{\alpha}$
separately in each sweep and thus obtain the averages $\rho$
for each sweep.  In Fig. \ref{rho-T}, we show the average spin
stiffness as a function of the temperature for system sizes $L=12,24,
\dots 384$. The steep drop in $\rho$ which occurs as $T$
increases beyond $T_{th} \simeq 0.28$ is consistent with the rapid
decrease of $\xi$ in the same temperature regime in which the
vortices become unbound \cite{KM84}. In contrast to the spin stiffness
of the planar XY model, $\rho$ does not saturate in the low
temperature regime $T<T_{th}$ with increasing system size $L$ but
decreases with increasing $L$. This behavior is to be expected, since
in contrast with the correlation length of the XY model, the
correlation length of the HAFT remains finite for low temperatures,
where the vortices are bound in pairs. As $T \rightarrow 0$,
our data for $\rho$ approach the correct limiting value $1/\sqrt3$.

In Fig. \ref{rho-L}, we display $\rho$ for various
temperatures and system sizes. The data are averages over 3-5
independent runs of lengths comparable to those which we have
described above, error bars indicate the standard deviation.

In the low temperature regime, the thermodynamics of the classical
HAFT should be captured by the appropriate NL$\sigma$ model
\cite{DR89,ADM92,AWE92}. On the basis of a renormalization group
treatment of this model, Azaria {\it et al.} \cite{ADJM92} have made
detailed predictions for the dependence of the spin stiffness on the
linear system size $L$ and the correlation length $\xi$.  In their MC
study of the classical HAFT, Southern and Young \cite{SY93} found
excellent agreement with the predicted $L$ dependence of the spin
stiffness tensor at the temperature $T=0.2$.

In order to be able to compare our numerical results with the
predictions of the RG analysis of the NL$\sigma$ model, we integrated
the two loop RG equations of Azaria {\it et al.} \cite{ADJM92}
starting from the initial conditions $\rho (L=\Delta)=1/\sqrt3$
and $\rho _3(L=\Delta)/\rho _1(L=\Delta)=2$. Here, $\rho _1$ and $\rho
_3$ are the two main components of the spin stiffness tensor with
respect to the reference frame of the local order parameter
\cite{SY93}. By the above initial conditions we identify $\rho _1$ and
$\rho _3$ with their microscopic values on the scale of the lattice
constant $\Delta$. We find that in the temperature regime under
consideration, $T\leq 0.3$, the average stiffness $\rho$
varies linearly with $\ln L$ to a very good approximation on the scale
$12\leq L\leq 384$ covered by our simulations, see Fig.
\ref{rho-lnksi} below.
{}From the RG equations, one can also infer that the slope
$\rho '(L)=-d\rho (L)/d\ln L$ decreases from
$\rho '=T/(3\pi)$ to $\rho '=T/(4\pi)$, when $L$ increases from
a value of the order of the lattice constant, $L \sim \Delta$, to
a value of the order of the correlation length, $L \sim \xi $. In
Fig. \ref{rho-L}, the straight lines are least squares fits to the
data. The slopes of these lines, normalized to the maximal theoretical
slope $\rho '=T/(3\pi)$, are tabulated in Table \ref{Gefaelle}.
For $T\leq 0.25$, the slopes are seen to be rather close to their
maximal value which obtains, when $L$ is of the order of the lattice
constant. This is not unexpected since according to the RG
calculations, the correlation length is many orders of magnitude
larger than our maximal system size $L=384$ for these temperatures.
The larger values of $\rho '$ which we find for $T\geq 0.28$ are
incompatible with the RG theory. Hence we conclude that for $T\geq
0.28$ unbound vortices, not being taken into account by the RG
analysis, begin to limit the range of the spin correlations in
the HAFT.

If one defines the correlation length $\xi$ through the matching
condition $\rho (L=\xi)=0$, then the result of the integration of
the two loop RG equations can be cast into the following form
\begin{equation}
\rho = f(T,L) \; \ln (\xi/L) \;\; . \label{10}
\end{equation}
This is shown for two different temperatures in the two graphs in Fig.
\ref{rho-lnksi}. Obviously, the function $f(T,L)$ depends weakly on L.
The relation Eq. (\ref{10}) makes it possible to obtain the correlation
length from the Monte Carlo simulation, even in the low temperature
regime where $\xi$ is much larger than the system size $L$. Inserting
our MC data for $\rho $ into Eq. (\ref{10}) and solving for $\xi$, we
obtain the data points shown in Fig. \ref{Reklame} for $T \leq 0.28$.
This Figure also includes the data for $T \geq 0.28$ which have
already been shown in Fig. \ref{ksi-T}.
The dashed and solid lines in Fig. \ref{Reklame} represent fits of
the RG and KT forms, Eqs. (1) and (2), to the MC data for $T \leq 0.28$
and $T \geq 0.28$, respectively.

\section{Summary}
We have shown that the MC simulation of the HAFT yields compelling
evidence for the influence of vortices on the spin correlations and
hence on the thermodynamics of this model. In agreement with earlier
findings by Kawamura and Miyashita \cite{KM84} we find that the
disordering effect of the vortices sets in rather abruptly at a
temperature $T_{th} \simeq 0.28$. Up to this temperature, our data
for the spin stiffness $\rho$ and the ensuing temperature dependence
of the correlation length $\xi$ are consistent with the predictions of
the RG analysis of the HAFT which ignores the existence of topological
defects such as vortices. For $T>T_{th}$, however, the simulation yields
temperature dependences of the correlation length $\xi$ and the
antiferromagnetic susceptibility $\chi({\bf Q})$ which are incompatible
with the RG predictions. Instead, in this temperature regime, the
temperature dependences of $\xi$ and $\chi({\bf Q})$ which follow from
the KT picture of unbinding vortex pairs provide satisfactory fits of
the data.
A rapid increase of the density of unbound vortices for $T>0.3$
had already been found by Kawamura and Miyashita in their simulation
of the HAFT \cite{KM84}. It had not been clear, however, whether this
phenomenon would lead to the same temperature dependences of the
correlations of the HAFT as had been predicted for the planar XY model
by Kosterlitz and Thouless \cite{KT73}. In our recent analytical study
\cite{WEA94}, we were led to the conclusion that this should indeed be
the case. The present numerical study fully supports this conclusion.
As we have discussed in Ref. \cite{WEA94}, the crossover transition from
the RG type behavior to the KT type behavior of the correlations of
the HAFT cannot imply a phase transition in the proper sense,
because the correlation length is finite both below and above the
transition temperature $T_{th}$. The results shown in Fig. \ref{Reklame}
indicate, however, that the transition happens in a very narrow
interval around $T_{th}$. Therefore we consider it possible that the
derivative of the correlation length with respect to the temperature
develops a discontinuity at $T_{th}$ in the thermodynamic limit.

\subsection*{Acknowledgement}
The numerical calculations were carried out in part at the Regionales
Rechenzentrum Niedersachsen, Hannover.

\begin{table}
\caption[]{\label{tab_ksi_chi}
   Correlation length $\xi$ and antiferromagnetic susceptibility
   $\chi({\bf Q})$ for system  sizes $L=96$, $192$, and $384$ for
   various temperatures.
   The statistical error represents the standard deviation over 3-5
   independent runs. The two last columns contain the RG predictions
   for $\xi$ and $\chi({\bf Q})$.
   Asterisks indicate agreement of the results for different system
   sizes, see main text. }

\begin{tabular}{rr|rrr|rr}
T & L & & $\xi$ & $\chi$ & $\xi_{RG}$ & $\chi_{RG}$ \\ \hline
0.300 & 384 &  &  105.0 $\pm$ 11.9 &   5014 $\pm$  543 &
\raisebox{-2ex}[0mm][2ex]{82.8} &   \raisebox{-2ex}[0mm][2ex]{3609} \\
0.300 & 192 &  &   83.5 $\pm$  3.5 &   2800 $\pm$  135 &    &    \\ \hline
0.305 & 384 & \raisebox{-2ex}[0mm][2ex]{*} &   60.7 $\pm$  6.7 &   2184 $\pm$
273 &   \raisebox{-2ex}[0mm][2ex]{57.0} &   \raisebox{-2ex}[0mm][2ex]{1796} \\
0.305 & 192 &  &   63.4 $\pm$  3.8 &   1899 $\pm$   55 &    &    \\ \hline
0.310 & 192 &  &   41.3 $\pm$  1.8 &   1040 $\pm$   39 &
\raisebox{-2ex}[0mm][2ex]{39.7} &    \raisebox{-2ex}[0mm][2ex]{914} \\
0.310 &  96 &  &   38.2 $\pm$  0.3 &    748 $\pm$   20 &    &     \\ \hline
0.315 & 192 &  &   31.7 $\pm$  1.1 &    594 $\pm$   25 &
\raisebox{-2ex}[0mm][2ex]{28.0} &    \raisebox{-2ex}[0mm][2ex]{476} \\
0.315 &  96 &  &   29.3 $\pm$  0.7 &    494 $\pm$   11 &    &     \\ \hline
0.320 & 192 & \raisebox{-2ex}[0mm][2ex]{*} &   22.7 $\pm$  2.0 &    349 $\pm$
27 &   \raisebox{-2ex}[0mm][2ex]{19.9} &    \raisebox{-2ex}[0mm][2ex]{253} \\
0.320 &  96 &  &   22.6 $\pm$  1.1 &    350 $\pm$   13 &    &     \\ \hline
0.330 & 192 & \raisebox{-2ex}[0mm][2ex]{*} &   14.3 $\pm$  3.4 &    169 $\pm$
17 &   \raisebox{-2ex}[0mm][2ex]{10.4} &     \raisebox{-2ex}[0mm][2ex]{76} \\
0.330 &  96 &  &   14.5 $\pm$  1.5 &    169 $\pm$    9 &    &      \\ \hline
0.340 & 192 & \raisebox{-2ex}[0mm][2ex]{*} &   11.8 $\pm$  3.3 &    103 $\pm$
 7 &    \raisebox{-2ex}[0mm][2ex]{5.7} &     \raisebox{-2ex}[0mm][2ex]{25} \\
0.340 &  96 &  &    9.8 $\pm$  1.1 &     96 $\pm$    5 &     &      \\
\end{tabular}
\end{table}
\begin{table}
\caption[]{\label{Gefaelle}
   Slopes of the fits in Fig. \ref{rho-lnksi} normalized to the
   maximum slope $T/(3\pi)\\$.}

\begin{tabular}{r|rrrrrr}
T & 0.10 & 0.20 & 0.25 & 0.28 & 0.29 & 0.30 \\
$\bar{\rho}'/\bar{\rho}'_{max}$& 0.962& 0.937& 0.939& 1.045& 1.191& 2.098 \\
\end{tabular}
\end{table}

\begin{figure}
\caption[]{ \label{ksi-T}
   Correlation length $\xi$ as a function of the temperature $T$.
   Solid line: the KT form Eq. (\ref{2}) with $C^{\xi}_{KT}=0.47$,
   $b=0.77$, and $T_{th}=0.28$.
   Dashed line: the RG behavior Eq. (\ref{1}) with
   $C^{\xi}_{RG}=3 \cdot 10^{-8}$.
}
\end{figure}
\begin{figure}
\caption[]{\label{chi-T}
   Antiferromagnetic susceptibility $\chi({\bf Q})$ as a function of the
   temperature $T$.
   Solid line: the KT form Eq. (\ref{7}) with $C^{\chi}_{KT}=0.40$,
   $b=0.77$, and $T_{th}=0.28$.
   Dashed line: the RG behavior Eq. (\ref{6}) with
   $C^{\chi}_{RG}=6 \cdot 10^{-12}$.
}
\end{figure}
\begin{figure}
\caption[]{\label{ksi-wT}
   Correlation length $\xi$ as a function of $(T-T_{th})^{-\frac{1}{2}}$.
   The lines are defined as in Fig. \ref{ksi-T}.
}
\end{figure}
\begin{figure}
\caption[]{\label{chi-wT}
   Antiferromagnetic susceptibility
   $\chi({\bf Q})$ as a function of $(T-T_{th})^{-\frac{1}{2}}$.
   The lines are defined as in Fig. \ref{chi-T}.
}
\end{figure}
\begin{figure}
\caption[]{\label{rho-T}
   Spin stiffness $\rho $ as a function of the temperature $T$.}
\end{figure}
\begin{figure}
\caption[]{\label{rho-L}
   Spin stiffness $\rho $ as a function of the system size $L$.
   The straight lines are mean squares fits, their slopes are tabulated in
   Tab. \ref{Gefaelle}.
}
\end{figure}
\begin{figure}
\caption[]{\label{rho-lnksi}
   Spin stiffness $\rho $ as a function of $\ln (\xi/L)$ as obtained from the
   two loop RG equations. The curves correspond to $T=0.1$ and $0.25$,
   respectively.
}
\end{figure}
\begin{figure}
\caption[]{\label{Reklame}
   Correlation length as a function of the temperature. Dashed lines:
   fit of the RG result to the low temperature data. Solid line:
   fit of the KT form to the high temperature data.
   The inset shows the data on a larger temperature scale.
}
\end{figure}


\begin{thebibliography}{10}

\bibitem{EXP90}
H. Kadowaki, H. Kikuchi, and Y. Ajiro, J. Phys. C: Solid State Phys. {\bf 2},
  4485  (1990);
K. Takeda, K.Miyake, K. Takeda, and H. Hirakawa, J. Phys. Soc. Jpn. {\bf 61},
  2156  (1992;
N. Kojima {\it et~al.}, J. Phys. Soc. Jpn. {\bf 62},  4137  (1993).

\bibitem{ADJ90}
P. Azaria, B. Delamotte, and T. Jolicoeur, Phys. Rev. Lett. {\bf 64},  3175
  (1990).

\bibitem{ADM92}
P. Azaria, B. Delamotte, and D. Mouhanna, Phys. Rev. Lett. {\bf 68},  1762
  (1992).

\bibitem{AWE92}
W. Apel, M. Wintel, and H.~U. Everts, Z. Phys. B - Condensed Matter {\bf 86},
  139  (1992).

\bibitem{LR93}
P. Leung and K. Runge, Phys. Rev. B {\bf 47},  5861  (1993).

\bibitem{DE93}
R. Deutscher and H.~U. Everts, Z. Phys. B - Condensed Matter {\bf 93},  77
  (1993).

\bibitem{KM84}
H. Kawamura and S. Miyashita, J. Phys. Soc. Jpn. {\bf 53},  4138  (1984).

\bibitem{M79}
N. Mermin, Rev. Mod. Phys. {\bf 51},  591  (1979).

\bibitem{KT73}
J.~M. Kosterlitz and D.~J. Thouless, J. Phys. C: Solid State Phys. {\bf 6},
  1181  (1973).

\bibitem{KK93}
H. Kawamura and M. Kikuchi, Phys. Rev. B {\bf 47},  1134  (1993).

\bibitem{WEA94}
M. Wintel, H.~U. Everts, and W. Apel, Europhys. Lett. {\bf 25},  711  (1994).

\bibitem{SY93}
B.~W. Southern and A. Young, Phys. Rev. B {\bf 48},  13170  (1993).

\bibitem{CADM94}
M. Caffarel, P. Azaria, B. Delamotte, and D. Mouhanna, Europhys. Lett. {\bf
  26},  493  (1994).

\bibitem{DR89}
T. Dombre and N. Read, Phys. Rev. B {\bf 39},  6797  (1989).

\bibitem{ADJM92}
P. Azaria, B. Delamotte, T. Jolicoeur, and D. Mouhanna, Phys. Rev. B {\bf 45},
  12612  (1992).

\end{thebibliography}
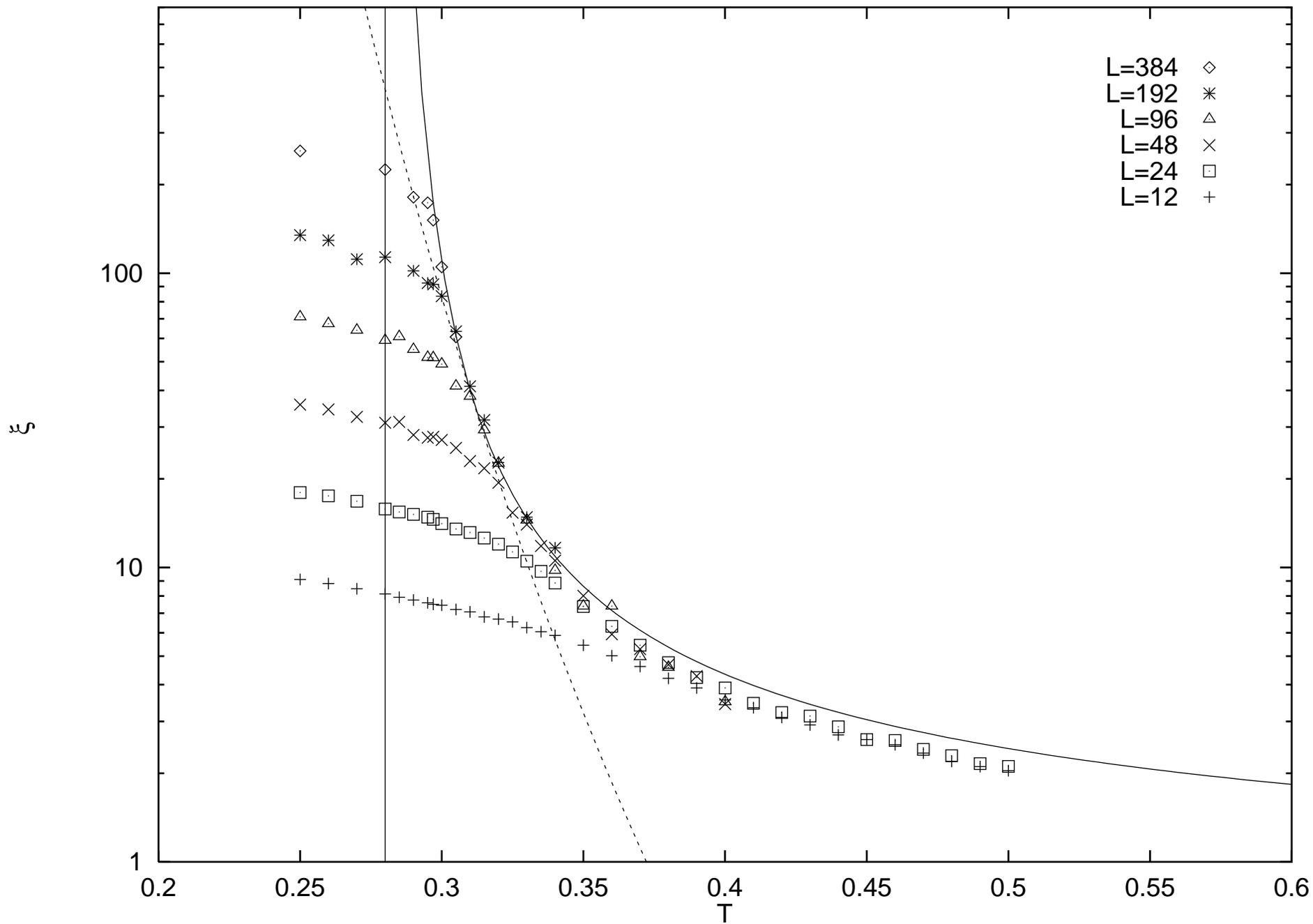
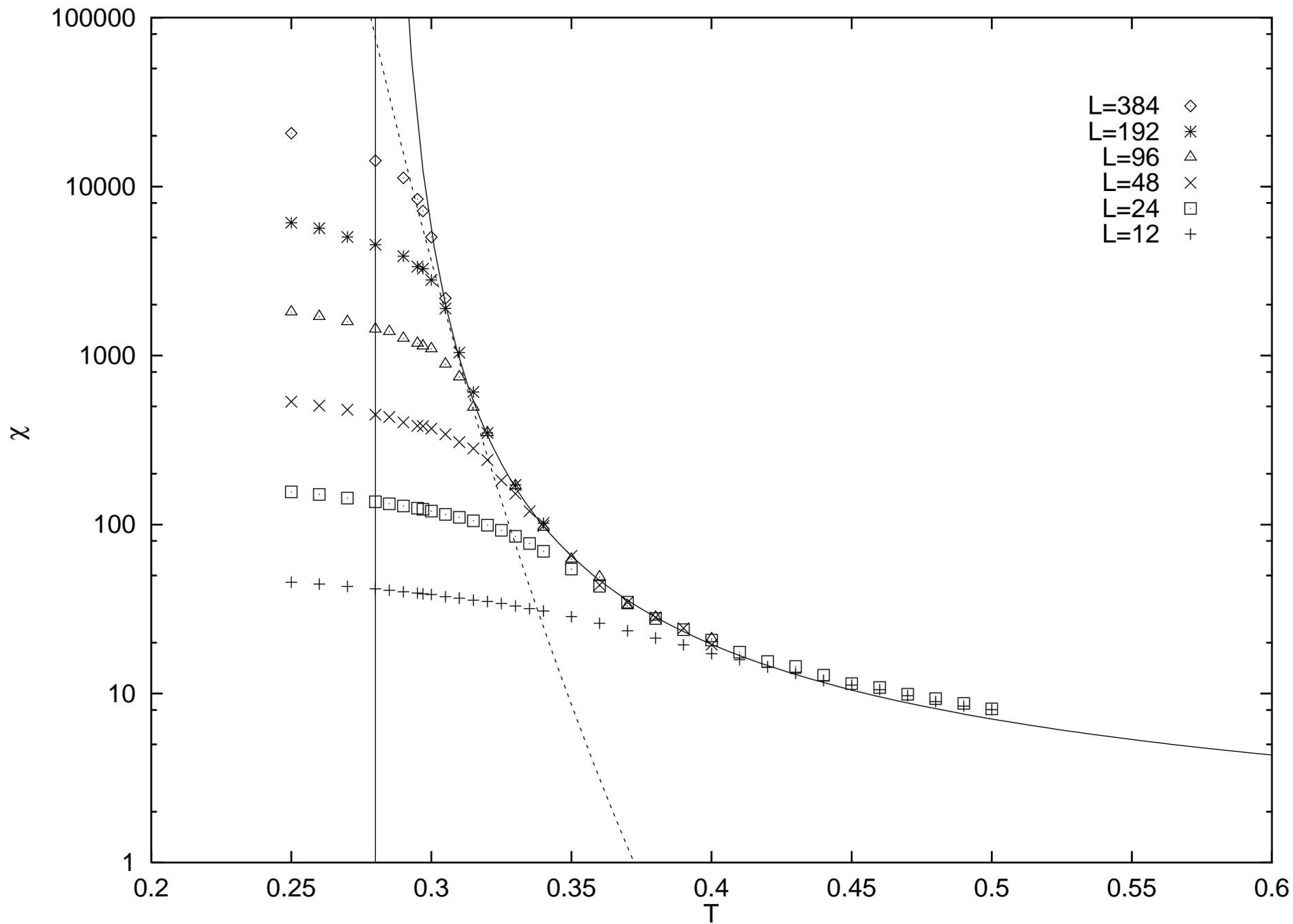
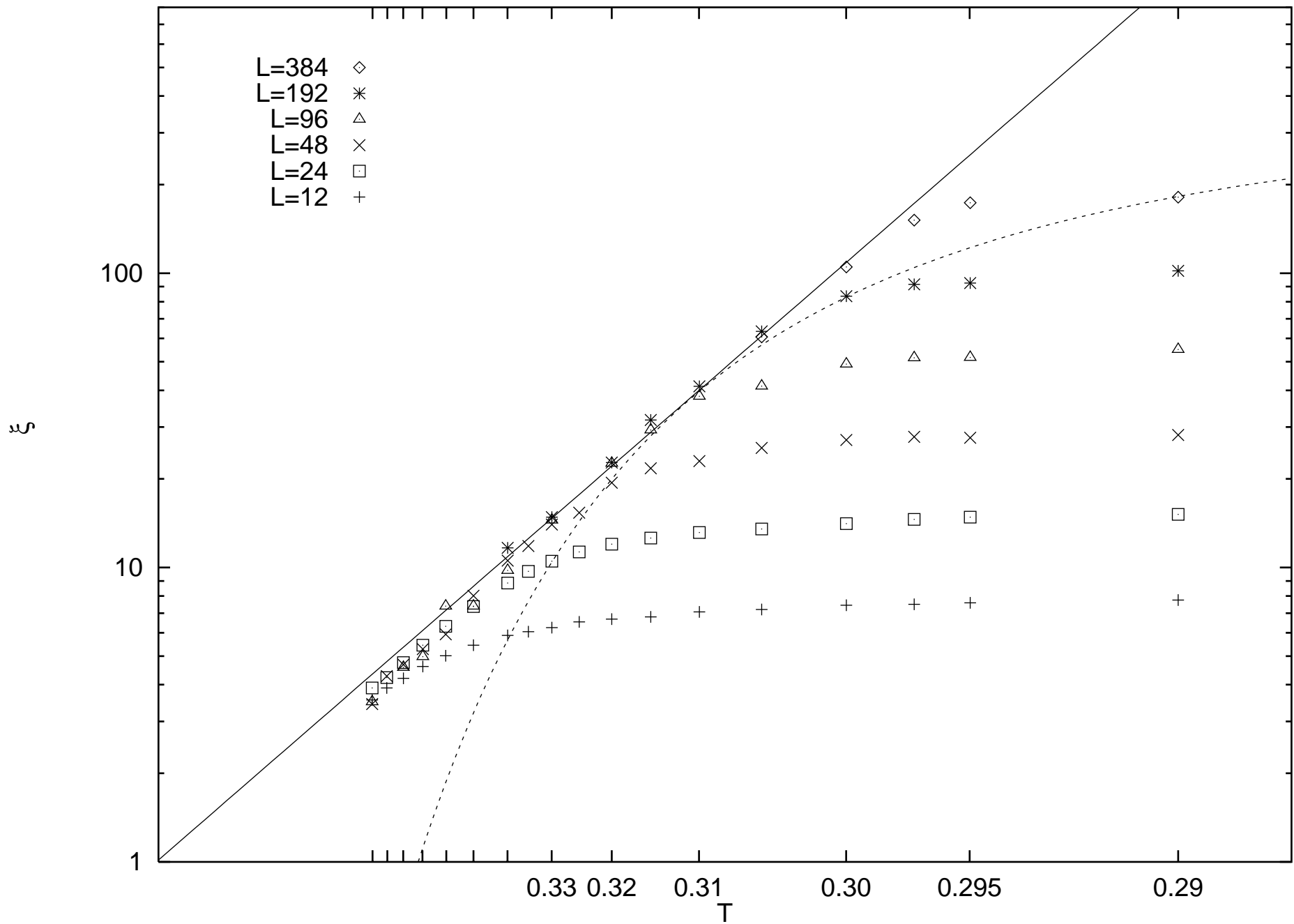
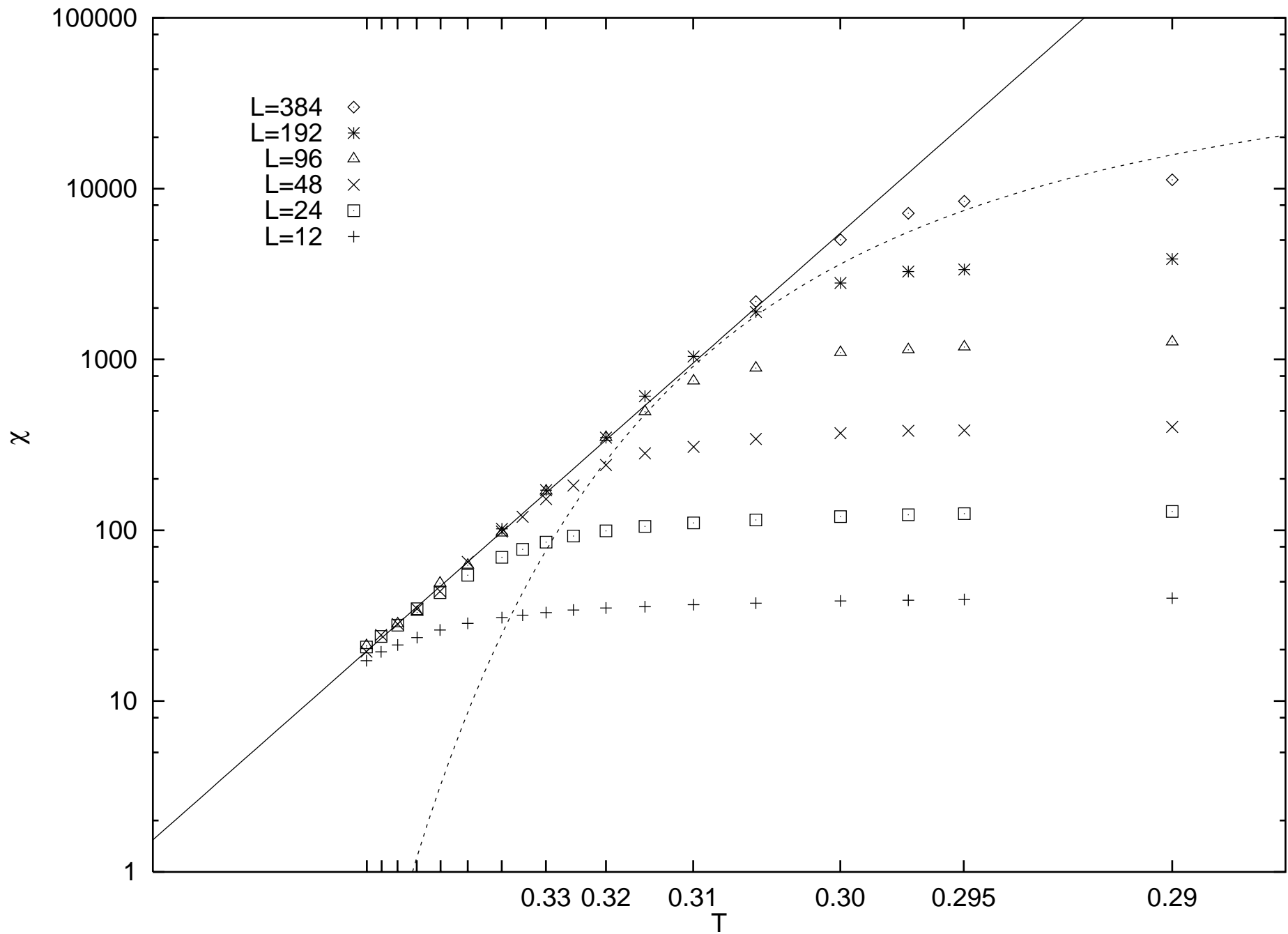
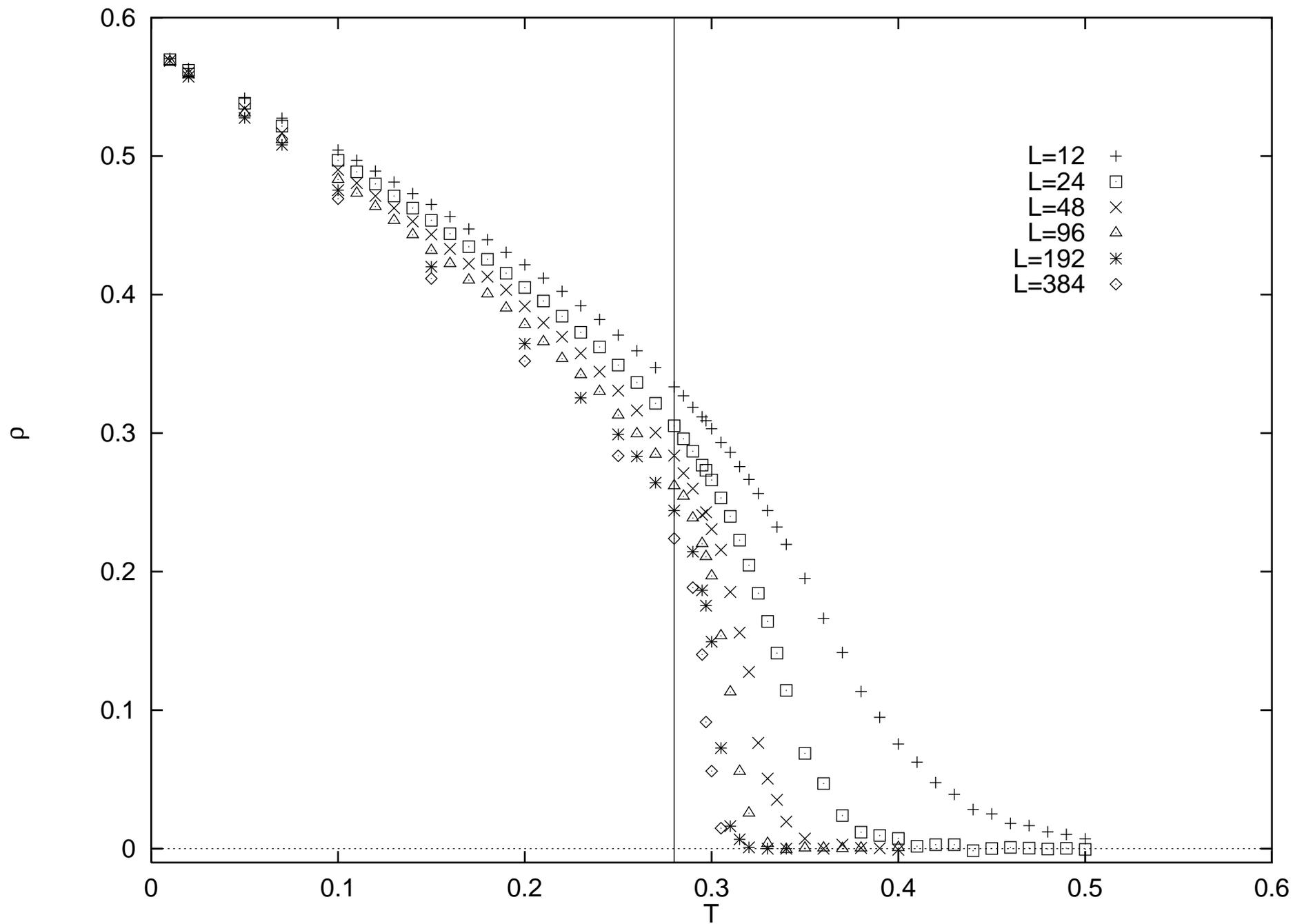
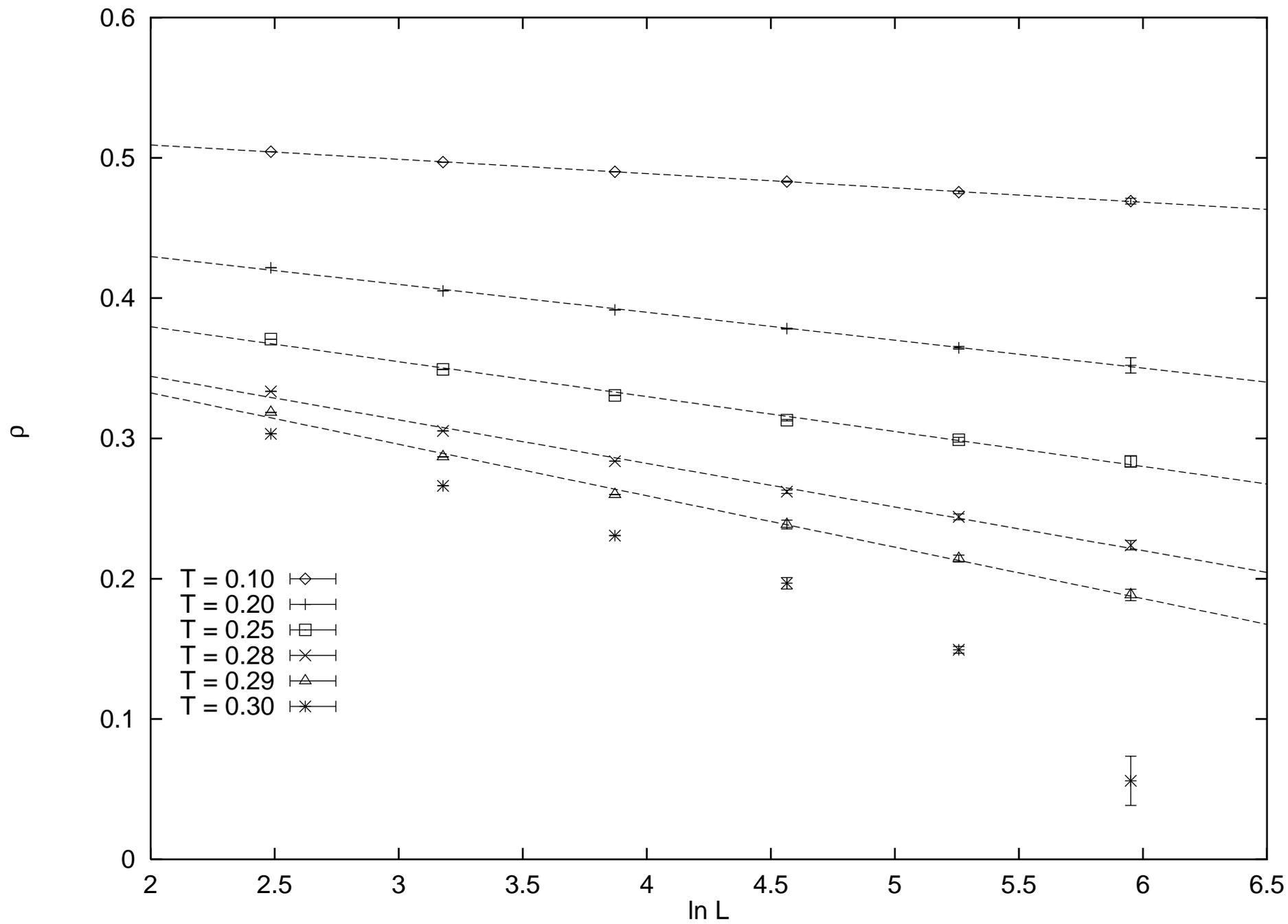
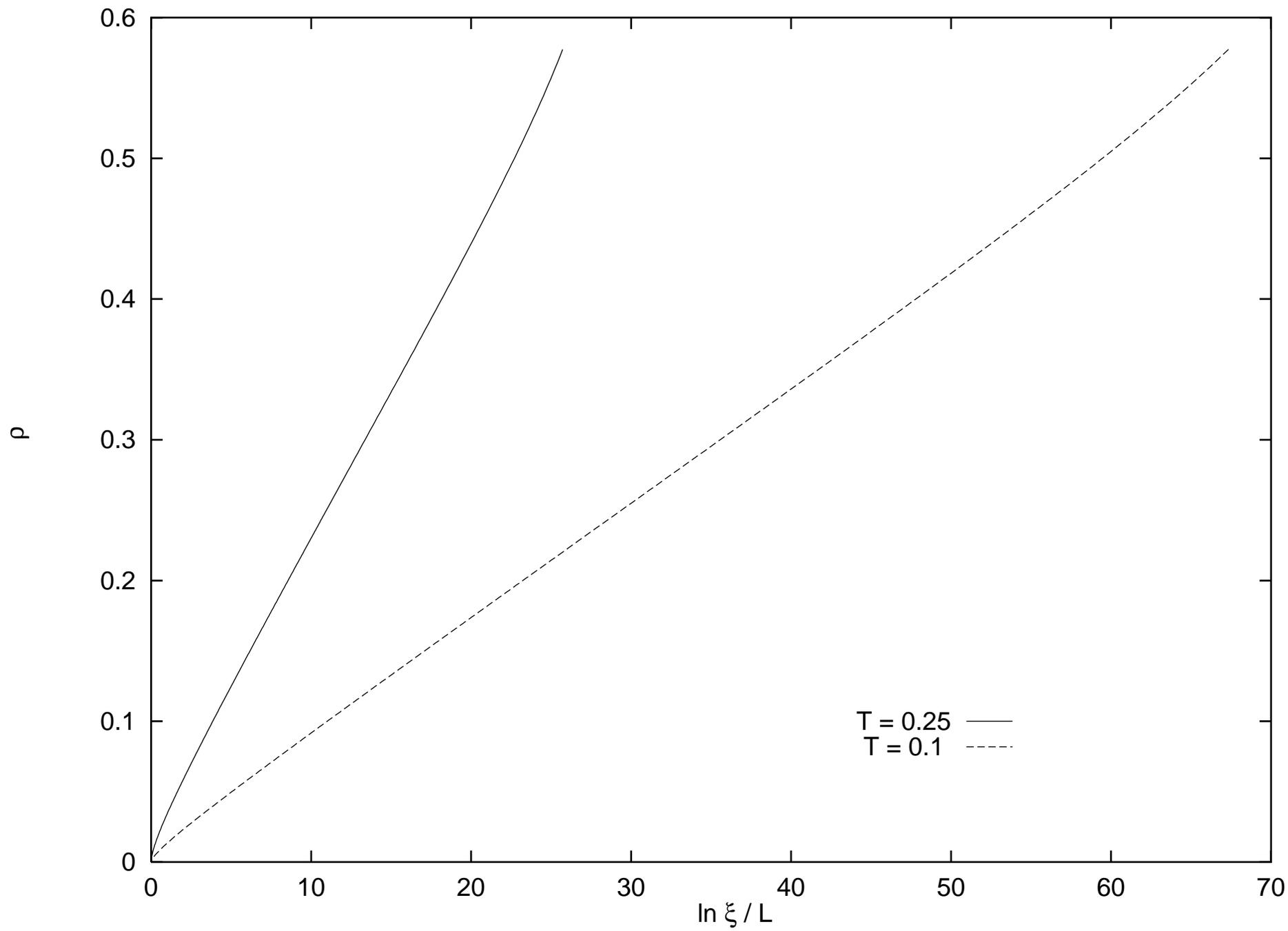
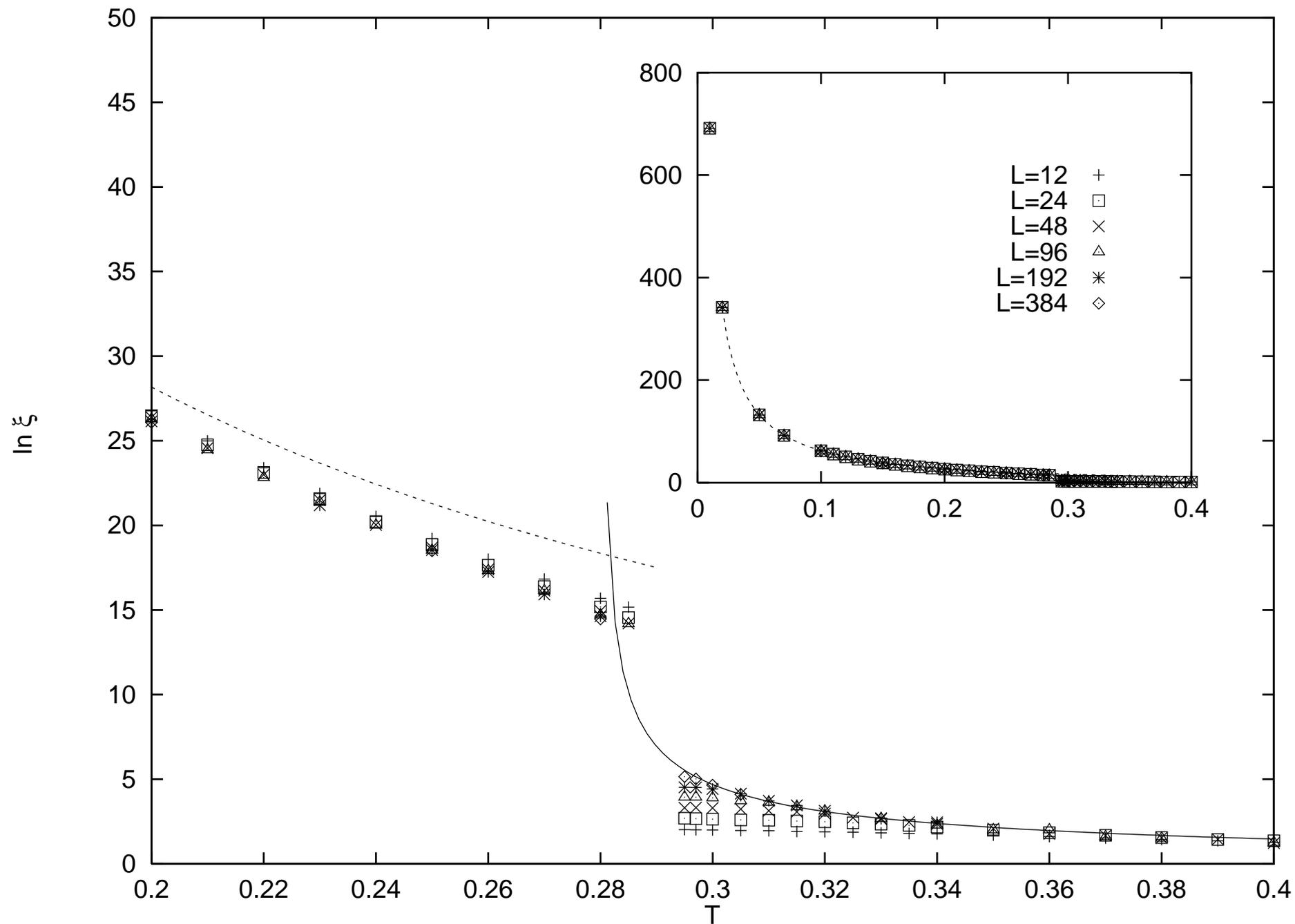
\end{document}